\documentclass{article}  
\usepackage{jamaica04}

\usepackage{graphicx}
\frompage{000} \topage{000}                                              
\def\rightmark{Coulomb Dissociation of $^8B$}
\def\leftmark{Moshe Gai}

\title{{\bf The Coulomb Dissociation of $^8B$  and a \\  
Critical Assessment of the Seattle $S_{17}(0)$ Result.
\footnote{Work Supported by USDOE Grant No. DE-FG02-94ER40870.}
}} 
\authors{
{Moshe Gai}\\
{\normalsize 
Laboratory for Nuclear Science at Avery Point, \\ University of 
Connecticut, 1084 Shennecossett Road, Groton, CT 06340. \\
gai@uconn.edu, URL: http://www.phys.uconn.edu.}}

\abstract{The Coulomb dissociation of $^8B$, measured with high precision by the GSI 
group, is in excellent agreement with the astrophysical cross section factor ($S_{17}$) 
measured by the Weizmann group with a $^7Be$ target. The GSI and Weizmann data are in 
good agreement with the Seattle data at high energies, but at low energies we 
observe a slight systematic (up to 2$\sigma$) deviation, yet the Seattle group repeatedly 
rejects the CD method. We show that when plotting the slopes, they mis plotted 
one CD data point and exclude measured slopes that contradict their claim. 
Indeed the measured slope is essential to elucidate the d-wave 
correction to $S_{17}(0)$ that could be as large as 15\%, and is the last open question 
that needs to be resolved before $S_{17}(0)$ can be quoted with an accuracy of 5\% or 
better. We show that this goal has not been achieved (in spite of the strong claim of 
the Seattle group), since  currently there is no agreement among experiments as well 
as among theoretical models on the value of the slope. In addition, currently there 
is no theoretical framework within which (for example the Seattle) data can be  
analyzed and $S_{17}(0)$ extrapolated with the claimed high precision of 2.5\%. This 
(last) issue of the slope and the d-wave correction must be resolved (by future measurements) 
so as to allow quoting $S_{17}(0)$ with an accuracy of 5\% or better.}

\begin{document}
 
\maketitle
\setcounter{page}{1}

\section{Introduction}\label{intro}

The astrophysical cross section factor of the $^7Be(p,\gamma)^8B$ reaction, $S_{17}(0)$
as defined in Ref. \cite{Adel}, is essential for predicting the $^8B$ solar neutrino flux 
\cite{SSM}, which is now measured with 7.3\% accuracy \cite{SNO}, and extracted from a global 
fit of solar and reactor neutrino experiments \cite{Bah03} with an accuracy of 4\%. Hence  
it is essential to know $S_{17}(0)$ with a comparable accuracy. While recent claims suggest 
\cite{Seatt} $S_{17}(0)$ measured with high accuracy, we demonstrate that this 
accuracy has not yet been achieved.

\section{Coulomb Dissociation}\label{Coul}

The Coulomb dissociation (CD) method \cite{Bauer} was applied to the 
dissociation of $^8B$. At first the RIKEN1 \cite{Mot94} and RIKEN2 \cite{Kik} data 
addressed the (historical) disagreement between data measured by Filippone and Vaughn 
\cite{Fil83,Vau70} and that of Kavanagh and Parker \cite{Parker,Kav69}. The  RIKEN1 
data were available seven years before any modern day measurement of the direct reaction. 
These new direct capture measurements \cite{Ham01,Str01} confirmed  
the finding of the RIKEN measurements that the lower value measured by 
Filippone and Vaughn is preferred. However, with the much improved accuracy 
a new problem arises vis-a-vis a disagreement between the Seattle data \cite{Seatt} 
that quoted cross section slightly larger than Filippone's with $S_{17}(0)$ 
approximately 22 eV-b, and a value slightly smaller than Filippone's of 
approximately 18.5 eV-b quoted by the Orsay \cite{Ham01} and the Bochum \cite{Str01} groups. Again, long before the new 
disagreement was even known and four years before the Weizmann group \cite{Weiz} 
confirmed the Seattle larger value, the high precision measurement of the CD of $^8B$ by 
the GSI1 group \cite{Iw99} and later the GSI2 group \cite{Sch03} confirmed the larger value 
quoted by the Seattle group of approximately 21 eV-b.

\begin{figure}[htb]
\vspace*{-.3cm}
\hspace{2cm} \includegraphics[width=8cm]{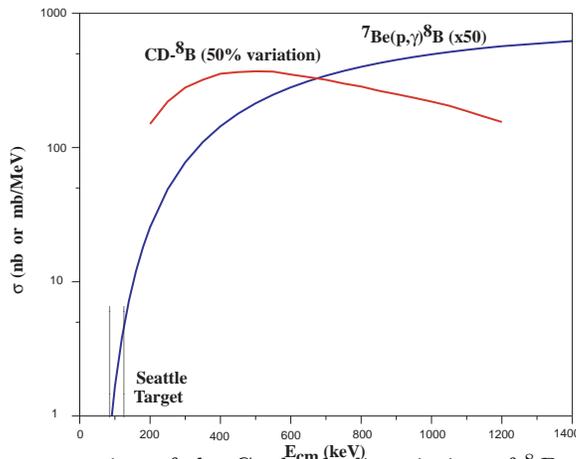}
\vspace*{-0.8cm}
\caption[]{The cross section of the Coulomb dissociation of $^8B$ as 
compared to the cross section of the $^7Be(p,\gamma)^8B$ direct reaction.}
\label{fig1}
\end{figure}

The Coulomb dissociation method is particularly useful for measuring absolute value of the 
cross section since the yield of the CD reaction is almost constant in the energy 
range of 200 - 1,300 keV, as shown in Fig. 1. In comparison the direct cross
section varies by a factor 50 over the same energy region. Note that it varies by a factor 
of five across the Seattle target, at the lowest measured energy of 112 keV \cite{Seatt}. 
But the CD yield varies by no more than 50\% over the energy range of 200 - 1,300 keV. In 
addition the CD method is ideal for measuring the slope (S' = dS/dE) of the cross section factor, 
as it is dependent mostly on a well understood virtual photon flux theory. 
The measured slope in direct 
capture measurement is dependent on several experimental parameters that need to 
be measured with high precision, not the least of which is the energy loss and dE/dX across 
the target. 

Note that Filippone's \cite{Fil83} target is three times thinner than Seattle's \cite{Seatt}. It 
was studied in detail (and published in a separate paper) with low energy 
proton resonance at 441.4 keV. Yet Filippone {\em et al.} 
conservatively quote an uncertainty of 11\% at the lowest measured energy, due to uncertainty in 
dE/dX across the target. Filippone's conservative attitude must be compared with the 2.5\% uncertainty 
\cite{Seatt} quoted at the same low energy point, with a target three times 
thicker, and energy loss that is studied at 1.4 MeV with alpha-particles and not protons.

\section{The Slope of $S_{17}$ and the d-Wave Correction}

The d-wave correction of $S_{17}(0)$, first introduced by Robertson \cite{Ham73}, was 
developed by Barker \cite{Barker} and estimated by Xu {\em et al.} \cite{Xu94} and 
Jennings {\em et al.} \cite{Jen}, and was shown by Filippone \cite{Fil83} 
to reduce $S_{17}(0)$ by as much as 15\%. The d-wave correction is directly 
related to the slope (S' = dS/dE) at energies above 300 keV \cite{Jen,Xu94}. Thus an 
accurate knowledge of the slope (S') is essential for an accurate knowledge of 
the d-wave correction to $S_{17}(0)$.
As we demonstrate below, the slope S' at energies above 300 keV is not very well known, 
hence the d-wave correction is ill determined. The d-wave correction of up to 15\% 
\cite{Fil83} is up to six times larger than the quoted theoretical uncertainty of 2.5\% 
\cite{Seatt}. For this and other reasons we doubt the validity of the impressive quoted 
theoretical accuracy of 2.5\% \cite{Seatt}.

\begin{figure}[htb]
\vspace*{-0.3cm}
\hspace{1cm} \includegraphics[width=10cm]{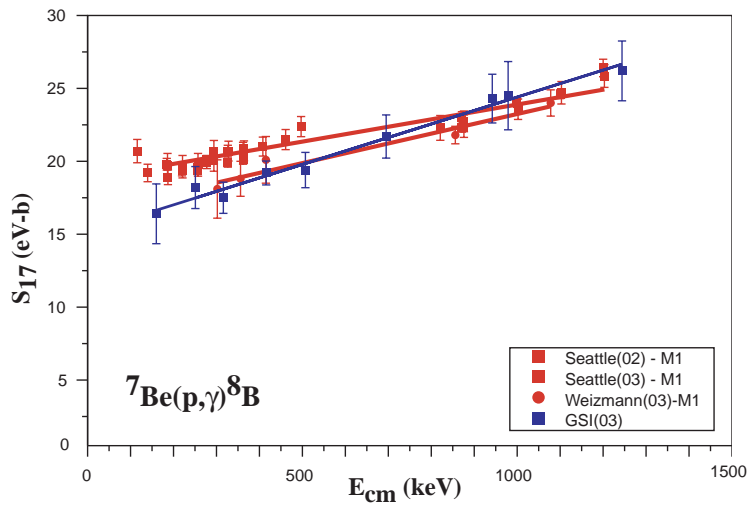}
\vspace*{-0.6cm}
\caption[]{A comparison of the Seattle, Weizmann, 
and GSI data.}
\label{fig2}
\vspace{-1cm}
\end{figure}

\section{Comparison of World Data}

The agreement between the Weizmann \cite{Weiz} and GSI \cite{Iw99,Sch03} data 
is excellent, as shown in Fig. 2. And the Weizmann and GSI data agree with 
Seattle data at higher energies, but are systematically smaller at lower energies as shown in 
Fig 2. The agreement between the GSI CD data and the direct capture reaction 
data of Weizmann (as well as Seattle) is in sharp contrast to the disagreement between the 
Seattle data \cite{Seatt} and the Bochum and Orsay data \cite{Ham01,Str01}. 
For example the Bochum data does not agree with a 
single data point measured by Seattle, and in general the disagreement among these
data is by three to five sigma.

The data of the Weizmann group exhibit a different slope than Seattle's as shown for example 
in Fig. 19 of the Phys. Rev. C publication of the Seattle group \cite{Seatt}, as does the 
slope of the GSI1 and GSI2 data. However, the Seattle group has mis plotted the slope of 
the RIKEN2 data \cite{Kik} in their Fig. 19 of \cite{Seatt}. We refer the reader 
to Fig. 4 of the GSI2 paper \cite{Sch03}, from 
which it is clear that the slope of the RIKEN2 data is smaller than that of the GSI data. 
In addition the Seattle group claimed an agreement on the slope among direct capture measurements 
as well as an agreement among CD data, and a  
disagreement between CD and the direct capture slopes. In their attempt to present an agreement on 
the slope measured in direct capture, they use the Bochum and Orsay data (that disagree by up to 
5$\sigma$ with the Seattle data), but ignore the data measured by Vaughn, Parker and 
Kavangh \cite{Vau70,Parker,Kav69}. The first two were deemed non-useful for extracting $S_{17}(0)$ 
by Adelberger {\em et al.} \cite{Adel} due to the lack of low energy data (below 400 keV). But the 
measurement of a slope requires data above 400 keV as measured by Vaughn, Parker and Kavanagh 
and these data should not be ignored.

When all available data are used to extract the slope b from a fit to S = a(1+bE), as shown 
in Fig. 3, we conclude that there is no consensus on the slope among direct capture data, nor 
can we suggest that all CD data measure the same slope, nor do we conclude that there is a 
clear disagreement between CD data and direct data. The conclusion of the Seattle group that 
the slope is measured in direct capture reaction with high precision of 4.5\% \cite{Seatt}, is yet  
another statement of over reaching precision that can not be justified by the available 
data.

\begin{figure}[htb]
\vspace*{-0.3cm}
\hspace{1cm} \includegraphics[width=11cm]{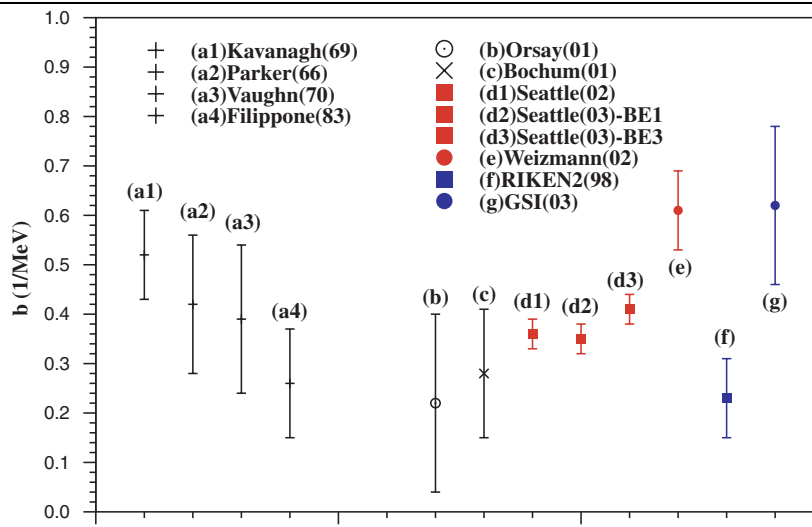}
\vspace*{-0.7cm}
\caption[]{Extracted slopes (b) from a fit to $S_{17}$ = a(1 + bE), 300 $<$ E $<$ 1400 keV. 
This figure correctly displays the slope of RIKEN2 data \cite{Kik} and it includes all available 
direct capture data, in sharp contrast to Fig. 19 of \cite{Seatt}.}
\label{fig3}
\vspace{-0.8cm}
\end{figure}

Furthermore, the very concept of extracting the "scale independent slope" (b) as 
opposed to S' = dS/dE = ab is a theoretical misconception.  Only at very low energies (below 
200 keV) where the reaction arises from an external capture, the logarithmic derivative 
S'/S(0) is an invariant \cite{Baye}. At higher energies the slope of the data 
is determined by the details of the nuclear potential and can not be assumed to have a simple 
relation to the ANC or the spectroscopic factor. This slope S' as we discussed above, is 
essential for elucidating the d-wave correction at zero energy and must be measured with 
high precision.

In Fig. 4 we compare the measured slopes (S') with theoretical predictions of approximately 
eleven theoretical papers including for example \cite{Sch03,Barker,Xu94,Jen,DB,DD}. 
The PRL data and the BE1 data included in the PRC article of the Seattle group 
\cite{Seatt} are outside the region predicted by theory. 
Indeed the so called agreement of the theoretical prediction of 
Descouvemont and Baye \cite{DB} with the slope of the Seattle data, was only due to the  
normalization (downward by a factor of 0.73) of the DB theory that changes the predicted slope by 
the same correction. The improved theory of Descouvemont and Douford 
\cite{DD} on the other hand requires a very small (5\%) adjustment (a normalization factor 
of 0.95 only), and the resulting slope does not agree with 
the Seattle data (with the best fit with a $\chi ^2 / \nu \ > \ 3$). 
In the absence of agreement between 
the improved theory and the Seattle data, we conclude that there is no theoretical frame work 
that can be used for extrapolating {\em with high precision} the Seattle data to zero energy. 
We emphasize again that a normalization of theoretical curves is required to 
compare the theory with data and for extrapolating to 
zero energy. But such a change of the slope amounts to a change of the predicted d-wave correction 
and hence it can not be used for {\em a precision extrapolation} of $S_{17}(0)$ (e.g. 2.5\%). Both 
the DD and Typel theory on the othre hand, fit the Weizmann and 
GSI data quite well with normalization factors of 
0.9 and 0.77, respectively, and yield $S_{17}(0)$ of 20.6 and 18.2, respectively, considerably 
smaller than the Seattle result and the value used in the SSM \cite{SSM}.

We conclude that a precision measurement of the slope (S') is required to 
allow extrapolating $S_{17}(0)$ with high precision of 5\% or better. It appears best to carry 
out this challenging measurement with $^7Be$ beams \cite{ISOLDE}.

\begin{figure}[htb]

\vspace*{-0.3cm}
\hspace{1cm} \includegraphics[width=9cm]{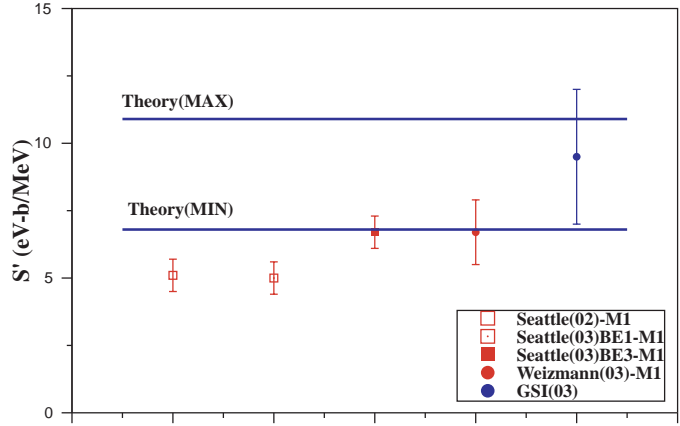}
\vspace*{-0.7cm}
\caption[]{Comparison of measured slopes (S' = dS/dE, 300 $<$ E $<$ 1400 keV)  
with predicted slopes. An M1 contribution due to resonance at 632 keV is 
subtracted from direct capture data.}
\label{fig4}
\vspace{-0.8cm}
\end{figure}

\vfill\eject
\end{document}